# Warming demands extensive tropical but minimal temperate management in plant-pollinator networks

Adrija Datta[1], Sarth Dubey[2], Tarik C. Gouhier[3], Auroop R. Ganguly[4,5,6], Udit Bhatia[2,7*]

[1]Department of Earth Sciences, Indian Institute of Technology, Gandhinagar, Gujarat, India-382355
[2]Department of Computer Science and Engineering, Indian Institute of Technology, Gandhinagar, Gujarat, India-382355
[3]Department of Marine and Environmental Sciences, Marine Science Center, Northeastern University, Nahant, MA 01908, USA
[4]Sustainability and Data Sciences Lab, Department of Civil and Environmental Engineering, Northeastern University, Boston, MA 02115, USA
[5]The Institute for Experiential AI and Roux Institute, Northeastern University, Boston, MA, USA
[6]Pacific Northwest National Laboratory, Richland, WA, USA
[7]Department of Civil Engineering, Indian Institute of Technology, Gandhinagar, Gujarat, India-382355
*Correspondence to: bhatia.u@iitgn.ac.in

**Abstract:**
Anthropogenic warming impacts ecological communities and disturbs species interactions, particularly in temperature-sensitive plant–pollinator networks. While previous assessments indicate that rising mean temperatures and shifting temporal variability universally elevate pollinator extinction risk, many studies often overlook how plant–pollinator networks of different ecoregions require distinct management approaches. Here, we integrate monthly near-surface temperature projections from various Shared Socioeconomic Pathways of CMIP6 Earth System Models with region-specific thermal-performance parameters to simulate population dynamics in 11 plant–pollinator networks across tropical, temperate, and Mediterranean ecosystems. Our results show that tropical networks, already near their thermal limits, face pronounced (~50%) pollinator declines under high-emissions scenarios (SSP5-8.5). Multi-species management targeting keystone plants emerges as a critical strategy for stabilizing these high-risk tropical systems, boosting both pollinator abundance and evenness. In contrast, temperate networks remain well below critical temperature thresholds, with minimal (~5%) pollinator declines and negligible gains from any intensive management strategy. These findings challenge single-species models and uniform-parameter frameworks, which consistently underestimate tropical vulnerability while overestimating temperate risk. We demonstrate that explicitly incorporating complex network interactions, region-specific thermal tolerances, and targeted multi-species interventions is vital for maintaining pollination services. By revealing when and where limited interventions suffice versus extensive management becomes indispensable, our study provides a clear blueprint for adaptive, ecosystem-specific management under accelerating climate change.

**Main Text:**

Biodiversity loss ranks among the most pressing global challenges[1] and has the potential to compromise the functioning of natural ecosystems as well as the services that they provide to growing human populations[2–4]. Climate change is projected to overtake habitat loss as the primary driver of biodiversity decline by the middle of the 21st century[5,6], posing unprecedented risks to ecosystems worldwide[7]. Shifts in species distributions and phenologies have already occurred in response to climate variability[8–10], signaling the wide-ranging effects of rising temperatures[11]. While targeted conservation actions can potentially mitigate biodiversity loss[12], the effectiveness of these measures hinges on our ability to predict how ecological systems will respond to environmental changes. Understanding these responses is crucial for mutualistic plant-pollinator networks, which underpin ecosystem health and food security but are increasingly vulnerable to warming[13].

Ectothermic insect pollinators are particularly sensitive to temperature changes[14–16], specifically in the tropics, where their narrow thermal niches can lead to heightened extinction risk[17]. This thermal sensitivity, which varies across ecoregions[18], suggests the need for region-specific management strategies[19], as uniform approaches may fail to address the distinct challenges posed by different ecological and climatic conditions. Although previous work has evaluated the impact of warming on the phenology and physiology of terrestrial ectotherms across latitudes[17,20], the underlying network structure and thermal thresholds of ecoregions were overlooked. Additionally, most ecological impact studies have relied on simplified models, such as single-species abundance-based approaches[17,20,21], to predict how climate change will affect ecological populations. While useful for identifying general trends, these models fall short of capturing the intricate, nonlinear interactions inherent in plant-pollinator networks because they rely solely on individual species' growth and carrying capacity parameters without accounting for interspecific interactions. Such interactions are vital for understanding the direct and indirect impacts of climate change on pollinator abundance, species diversity, and network stability across different climatic regions[22].

Additionally, studies that have incorporated region-specific network structure into their non-linear dynamical models have used fixed (mean) temperature differences to simulate the effects of climate change without accounting for shifts in its temporal variability across different ecoregions[23]. Furthermore, few studies have explored the potential benefits of multi-species management strategies compared to single-species approaches in promoting network persistence under different climate scenarios[23]. These limitations lead to considerable uncertainty regarding the effectiveness of current 'universal' management strategies for sustaining ecosystems around the globe, particularly since species' thermal properties vary across regions[24].

Tropical species often operate close to their thermal limits[17], making them more susceptible to even slight increases in temperature, while temperate regions typically remain below these critical thresholds, exhibiting less extreme responses to warming[17,20,24]. Such interspecific variability suggests that active management may produce different outcomes depending on the region. In addition, conservation efforts must weigh the trade-offs between enhancing mean species abundance while maintaining diversity and evenness. Strategies focused solely on boosting abundance may inadvertently reduce diversity[25], altering the balance of ecological interactions and potentially compromising long-term network stability. Additionally, not accounting for species' thermal responses when designing management interventions can lead to costly and ineffective strategies[26,27]. For instance, implementing intensive management in ecosystems that have yet to reach their thermal thresholds may yield minimal benefits, wasting resources and potentially destabilizing existing ecological balances[28]. Understanding whether multi-species management supports both abundance and diversity more effectively than single-species approaches is thus essential when tailoring strategies to the unique conditions of each ecoregion. Exploring these dimensions of ecosystem management under warming scenarios is crucial for designing conservation efforts that support biodiversity and overall ecosystem health efficiently.

In this study, we evaluate how different management strategies—specifically, single-species versus multi-species approaches—affect plant-pollinator networks (Fig. 1a-c) under warming scenarios in tropical, temperate, and Mediterranean ecosystems (Fig. 1d). We use data from the Web of Life database (https://www.web-of-life.es/), which details plant-pollinator interactions, with regional warming projections from Earth System Models (CMIP6) under historical, SSP2-4.5, and SSP5-8.5 scenarios in order to address two primary questions: (i) How do warming scenarios influence pollinator species abundance in distinct ecoregions? and (ii) Does a multi-keystone species management approach enhance overall abundance and diversity more than a single keystone species approach? Our findings reveal that ecosystems exhibit distinct responses to warming and management across regions: some benefit from active intervention as they approach critical thermal thresholds, while others require minimal measures to sustain stability. By incorporating multi-species interactions and network dynamics, we identify behaviors and thresholds that are overlooked in single-species analyses, challenging the notion that intensive management is universally required[29]. This work underscores the importance of region-specific strategies for targeted pollinator conservation under future climate scenarios, preserving biodiversity and ecosystem health where necessary.

*Regional variation in network structure and temperature dynamics*
Understanding how plant-pollinator networks respond to climate change requires a detailed analysis of their structural properties and the environmental factors that shape them. In this study, we assessed the impact of differential terrestrial warming in 11 plant-pollinator networks (Fig S1) across tropical, temperate, and Mediterranean regions, focusing on mean pollinator abundance and the implications for ecosystem health under different climate scenarios. These networks were selected to encompass a range of species richness (Fig S2) and topological differences (Fig 1a-c) e.g., connectance, asymmetry, nestedness (Fig S3), and climate conditions, which are crucial for understanding the broader effects of climate change on mutualistic interactions. We based the selection of these three ecoclimatic regions (Fig 1d) on the Köppen climate classification system, with temperate oceanic climate (Cfb) representing temperate regions, tropical savanna climate with a dry winter/dry summer (Aw) representing tropical regions, and hot summer Mediterranean climate (Csa) representing Mediterranean regions. This approach ensured the inclusion of region-specific biophysical parameters essential for our population models, a departure from earlier studies that often used constant parameters across diverse regions.

To explore how warming scenarios affect species' physiological parameters across ecoclimatic regions and influence pollinator abundance within networks, we utilized monthly near-surface air temperature (tas) data for historical (1850-2014) and future periods (2015-2100) under two warming scenarios (SSP2-4.5 and SSP5-8.5), sourced from ten Earth System Models (ESMs) (see Methods).

Our findings show that tropical regions maintain relatively stable year-round temperatures (Fig 1e, Fig S4a, Fig S5a), exhibiting minimal seasonal fluctuations (standard deviation: 1.28°C for historical, 1.39°C for SSP2-4.5, and 1.28°C for SSP5-8.5). However, future projections indicate these areas will face the most significant absolute temperature increases, particularly under SSP5-8.5 (28.62°C) and SSP2-4.5 (28°C). This suggests that despite their current stability, tropical networks are highly vulnerable to future warming due to their limited historical exposure to temperature variability. Conversely, temperate regions (Fig 1g, Fig S4c, Fig S5c) displayed pronounced seasonal temperature cycles (standard deviation: 4.59°C for historical, 4.80°C for SSP2-4.5, and 4.95°C for SSP5-8.5), with the Mediterranean region (Fig 1f, Fig S4b, Fig S5b) showing intermediate variability and marked summer warming. While projected temperature increases in temperate and Mediterranean regions are less extreme than in the tropics, the interaction of seasonal peaks and warming trends poses potential stress periods that could affect pollinator abundance and network stability. These results underscore the variability in thermal changes across ecoregions and the importance of developing management strategies tailored to local climatic and network structural characteristics. Such regional differences suggest that developing a 'one-size-fits-all' conservation strategy is unlikely to be successful. Accounting for these region-specific patterns of thermal variability is likely to be essential for developing effective conservation strategies to address the unique challenges posed by climate change to different ecosystems.

### *Differential warming and its impact on pollinator populations*

Our analysis of projected temperature trends indicates that all regions are expected to experience accelerated warming, with the tropical region (Fig 2a, Fig S6a) showing significant temperature increases (slope = 0.04, $p < 0.05$) under the SSP5-8.5 scenario. Temperate and Mediterranean regions (Fig 2b,c, Fig S6b,c) are also expected to experience increased warming, but the absolute temperature will remain lower than that of the tropical regions. The tropical region is expected to undergo a marked decline in pollinator species abundance (slope = -0.55, $p < 0.05$) (Fig 2d, Fig S7a), as are the Mediterranean (slope = -0.51, $p < 0.05$) (Fig 2e, Fig S7b) and the temperate regions (slope = -0.05, $p < 0.05$) (Fig 2f, Fig S7c). Under the SSP2-4.5 scenario, declines in population abundance are slower across all regions, but the trend persists.

It is important to note that the temperate region demonstrates relative mild declines in pollinator species abundance under both climate scenarios (SSP5-8.5: slope = -0.05, $p < 0.05$, Fig 2f, SSP2-4.5: slope = -0.05, $p < 0.05$, Fig S7c). In contrast, tropical and Mediterranean regions display more pronounced population declines. This trend is attributed to tropical and Mediterranean regions exceeding critical thresholds for various biophysical parameters, including growth and decay rates and competition strength. Additionally, these regions experience extended periods of temperatures above optimal levels, compounded by minimal seasonal variation, which limits recovery periods for species. Conversely, temperate regions also face warming but have yet to reach these critical biophysical thresholds. Temperature extremes in temperate areas are typically short-spelled, providing buffer periods that facilitate recovery and contribute to the slower observed decline in pollinator populations. These findings highlight the need for region-specific, adaptive conservation strategies that the unique dynamics of each ecoregion.

**Management strategies and their efficacy across ecoregions**
We analyzed the impact of the most extreme warming scenario (SSP5-8.5) on plant-pollinator networks under different management strategies—multi-pollinator, single-pollinator, multi-plant, single-plant management—and a no-management baseline from 2015 to 2100. The results revealed that the Management Efficiency Ratio (MER), defined as the ratio of mean annual pollinator abundance under a given management strategy to that observed without management, varies significantly across ecoregions, underscoring the region-specific nature of management effectiveness.

In tropical networks, management strategies yield limited benefits (Fig 3a, Fig S8a) when mutualistic strength is high ($\gamma_0 \geq 2$), but their influence grows as mutualistic strength decreases. Multi-species approaches, particularly those involving multiple plants or pollinators, show the greatest positive impact under moderate to low mutualistic strength ($0.001 < \gamma_0 \leq 1.5$), with effects becoming more evident after 2030. Mediterranean (Fig 3b, Fig S8b) and temperate (Fig 3c, Fig S8c) networks respond differently. Here, management interventions have limited impact unless mutualistic strength is extremely low ($\gamma_0 = 0.001$), where targeted multi-pollinator strategies become effective. This suggests that extensive management is only necessary under severe environmental conditions and when mutualistic interactions are relatively weak.

Interestingly, tropical networks exhibit shifts in effective management strategies based on mutualistic strength (Fig 3a, Fig S8a). Multi-plant management is most effective under moderate conditions, supporting connected pollinators, while multi-pollinator management becomes crucial under extremely weak mutualism ($\gamma_0 = 0.001$). In temperate regions (Fig 3c, Fig S8c), management remains non-essential mainly due to current biophysical resilience but may become relevant as these ecosystems approach critical thresholds toward the end of the century.

Mediterranean networks (Fig 3b, Fig S8b) demonstrate a mixed response, with targeted multi-pollinator management showing the highest efficacy only under the lowest mutualistic strengths. These findings emphasize that region-specific, adaptive management is vital for maintaining plant-pollinator ecosystems facing intensified warming.

We have depicted the statistical significance (Fig S9) of *t*-tests (with Bonferroni correction) comparing the effectiveness of four management strategies—Single Plant, Multi Plant, Single Pollinator, and Multi Pollinator—against a "no management" baseline at three significance levels: 1% ($p < 0.01$), 5% ($0.01 \leq p < 0.05$), and 10% ($0.05 \leq p < 0.1$). The results clearly show that management strategies are predominantly significant in tropical and temperate networks, whereas Mediterranean networks exhibit non-significant outcomes except at very low mutualistic strength levels ($\gamma_0 = 0.001$).

***Species rank-abundance and evenness shifts under management strategies***

To analyze changes in pollinator abundance distribution, we plotted the rank-abundance curve (RAC) for mean pollinator abundance from 2015 to 2100 across various management scenarios. In tropical regions (Fig 4a, Fig S10a), the RAC without management showed a steep initial slope, indicating dominance by a few species (relative abundance of 20% for the most abundant species) and numerous rare species, signifying low evenness (Peilou evenness index = 0.72 at $\gamma_0$ = 3) (Fig 4d, Fig S11a). This pattern underscores the vulnerability of tropical networks, where reliance on dominant species can amplify the risk of cascading failures if key species decline. Implementing plant management strategies (single and multi-plant) produced a more gradual slope, indicating improved evenness (Pielou evenness index = 0.85 at $\gamma_0$ = 3) through broader support for pollinator populations connected to managed plants. This suggests that plant-focused management can buffer against species dominance and enhance network stability by fostering a more balanced distribution of resources.

In contrast, pollinator-focused management boosted targeted species' abundance but maintained a steeper RAC, reflecting a network with lower evenness. This highlights a potential trade-off: while direct pollinator management can quickly increase population abundance for key species, it may not promote overall network stability or diversity.

Mediterranean (Fig 4b, Fig S10b) and temperate regions (Fig 4c, Fig S10c) displayed RACs with a gentler slope and shorter tail, signifying a more uniform species distribution (relative abundance of 4% for the most abundant species) and higher evenness (Pielou evenness index = 0.88 at $\gamma_0$ = 3) (Fig 4e,f, Fig S11b,c) without intervention. These findings suggest that such networks are inherently more balanced and potentially more resilient to environmental fluctuations, even under warming scenarios. Pielou's evenness index confirmed that higher mutualistic strength promoted evenness across all regions. However, tropical networks consistently lagged in evenness, indicating their susceptibility to shifts in species dynamics and environmental stressors.

Notably, tropical networks (Fig 4d, Fig S11a) benefited most from plant management, particularly multi-plant strategies, under weak mutualistic strength (evenness index difference of 0.36). This result implies that supporting plant diversity in highly interconnected and vulnerable ecosystems can cascade positively to pollinators, stabilizing the network. On the other hand, Mediterranean and temperate networks (Fig 4e,f, Fig S11b,c) exhibited minimal changes under any management, highlighting their reduced reliance on intervention and inherent stability.

These results collectively suggest that the efficacy of management strategies varies considerably across ecoregions. While targeted plant management can be transformative for tropical ecosystems nearing thermal thresholds, it may have limited impact in regions that naturally exhibit higher evenness and resilience. This suggests that universal conservation strategies may be ineffective and calls for adaptive, ecosystem-specific network approaches that simultaneously target species abundance and diversity.

**Conclusions**

The pursuit of universal patterns of resilience[30,31], tipping points[29,32,33] and strategies to mitigate biodiversity loss remains a critical goal in ecology and conservation. While progress has been made in defining early warning signals[34] and resilience[35,36] metrics, the divergent responses of ecosystems to climate change underscore the limitations of generalized approaches[37]. Each ecoregion's unique thermal thresholds[17] and biophysical characteristics[20] significantly influence stability and resilience, reinforcing the need for conservation strategies that align with regional ecological complexities.

Our findings demonstrate that plant-pollinator networks exhibit distinct responses to warming and management interventions depending on their ecological context. Tropical ecosystems, characterized by narrow thermal tolerances, show increased vulnerability and derive the most benefit from multi-species, habitat-focused management that supports both abundance and evenness. In contrast, Mediterranean and temperate regions exhibit more limited responses, indicating that intensive management might be less impactful or even unnecessary. These results challenge the prevailing view focusing on uniform, abundance-centric conservation strategies and advocate instead for adaptive, ecoregion-specific approaches that consider species diversity and overall ecosystem health. This shift is crucial in order to ensure that conservation efforts are not only effective but also avoid destabilizing ecosystems by intervening where natural resilience suffices.

While our network-based modeling approach provides essential insights, it comes with certain limitations. Expanding these models to include dynamic, time-dependent interactions would require comprehensive data on life stage-specific responses, thermoregulation[38], acclimatization[39], and adaptation—factors that are challenging to scale but could be crucial for accurately predicting ectotherm population dynamics. Furthermore, with their 1° spatial resolution, the climate projections employed may miss microclimatic variability[40], which can substantially influence localized responses and overall ecosystem resilience. Despite these limitations, our results provide key baseline expectations for both anticipating and effectively managing potential changes in pollinator abundance and diversity in response to warming across the globe.

We identified optimal management strategies for preserving species abundance, richness, and persistence by integrating region-specific pollinator networks with multi-model climate projections. Our results suggest that tropical ecosystems are most vulnerable, consistent with findings[17] but contrasting with variance-focused studies that indicate a higher risk in temperate regions[21,41]. This discrepancy illustrates how network-level interactions can reverse conclusions derived from single-species or variance-centric approaches. Critically, we demonstrate that tropical ecosystems benefit most from active multi-species management whereas temperate and Mediterranean regions remain comparatively stable with minimal intervention, challenging the notion of a one-size-fits-all strategy. These findings underscore the need for adaptive, region-specific conservation efforts that prioritize species diversity and ecosystem health informed by comprehensive climate projections and ecological models.

**Methods**

**Plant-Pollinator Networks**
We analyzed 11 bipartite plant-pollinator networks (Table S1) from three ecoclimatic regions—tropical, Mediterranean, and temperate—sourced from the Web of Life database (https://www.web-of-life.es/). Each network was represented as an unweighted, undirected bipartite graph, with nodes representing species (plants and insect pollinators) and edges denoting mutualistic interactions between species pairs. The selection of these networks was informed by the Köppen climate classification system[42]: temperate networks corresponded to the oceanic climate (Cfb), tropical networks to the savanna climate with dry winter/summer (Aw), and Mediterranean networks to the hot-summer Mediterranean climate (Csa). This classification aligned with the availability of region-specific biological parameters required for the mutualistic network model. Networks were chosen with 30 to 100 pollinator species to ensure representativeness across regions. For management interventions, single-species management involved numerical management of one species (plant or pollinator), whereas multi-species management targeted 10% of the total species (either plants or pollinators) in each network, chosen randomly to simulate the effects of broader ecosystem-level interventions.

**Future Climate Scenarios**

We obtained monthly near-surface air temperature data (tas) from 10 Earth System Models (ESMs) participating in the CMIP6 climate simulations: AWI-CM1.1-MR, BCC-CSM2-MR, CESM2-WACCM, CMCC-CM2-SR5, EC-Earth3, EC-Earth3-Veg, FGOALS-f3-L, INM-CM4-8, MRI-ESM2-0, and NorESM2-M. These data were sourced from the Earth System Grid Federation (https://esgf-node.llnl.gov/search/cmip6/) and covered both the historical period (1850–2014) and future climate scenarios under SSP2-4.5 and SSP5-8.5 pathways (2015–2100).

**Modelling temperature impacts on pollinators**

*Population dynamics modeling*

We incorporated interaction matrices from the selected networks into a mutualistic network model to simulate plant-pollinator dynamics. These models, governed by first-order differential equations (Eq. 1,2), capture the intricate population interactions within these communities. Insect pollinator abundance depends upon various biophysical parameters, i.e., intrinsic growth rate (α), decay rate(κ), intraspecific and interspecific competition(β), mutualistic strength(γ), handling time (*h*) and migration(μ). Mathematically, the population model is represented as follows:

$$\frac{\partial P_i}{\partial t} = P_i \left( \alpha_i^P(T) - \sum_{j=1}^n \beta_{ij}^P(T) P_j + \frac{\sum_{k=1}^m \gamma_{ik}^P A_k}{1 + h(T) \sum_{k=1}^m \gamma_{ik}^P A_k} \right) + \mu_P \quad (1)$$

$$\frac{\partial A_i}{\partial t} = A_i \left( \alpha_i^A(T) - \kappa_i^A(T) - \sum_{j=1}^m \beta_{ij}^A(T) A_j + \frac{\sum_{k=1}^n \gamma_{ik}^A P_k}{1 + h(T) \sum_{k=1}^n \gamma_{ik}^A P_k} \right) + \mu_A \quad (2)$$

In this modeling framework, $P_i$ and $A_i$ represent the abundances of the *i*-th plant and pollinator species, respectively, with $n$ and $m$ denoting the total number of plant and pollinator species within the network. The parameter $\alpha(T)$ describes the temperature-dependent intrinsic growth rate in the absence of competition and mutualistic effects, while $\beta_{ij}(T)$ represents temperature-dependent intraspecific and interspecific competition strength, respectively. $\kappa(T)$ denotes the temperature-dependent decay rate of that pollinator. The parameter $\mu$ denotes species immigration, and $h$ is the temperature-dependent handling time. The parameter $\gamma$ quantifies the strength of mutualistic interactions (Eq. 3). Generally, $\gamma$ depends on the degree, the number of mutualistic partners of species *i*, $D_i$, as follows:

$$\gamma_{ik} = \epsilon_{ik} \cdot \frac{\gamma_0}{D_i^t} \qquad (3)$$

where $\gamma_0$ is average mutualistic strength, $\epsilon_{ik} = 1$, if there is mutualistic interaction between species *i* and *k* or 0 otherwise, t modulates the trade-off between the interaction strength and the number of mutualistic links.

*Thermal influence on species' biophysical parameters*

Recent empirical studies demonstrated that species' biological parameters(e.g., growth rate $\alpha(T)$, decay rate $\kappa(T)$, handling time *h(T)* are functions of temperature[43]. These studies also investigated how constant temperature range (0°C to 40°C) and network structure affect stability criteria and identify tipping points[23]. Unlike previous studies, we utilized the mean monthly near-surface temperature (tas) for tropical, mediterranean and temperate regions for SSP2-4.5 and SSP5-8.5 to calculate temperature-dependent biological rates and hence calculated population abundance.

We considered temperature-dependent species' growth rate $\alpha(T)$ exhibiting a unimodal symmetric response (Eq. 4) represented by a Gaussian function[20,44].

$$\alpha_i(T) = \alpha_{opt} \cdot e^{-\frac{(T-T_0)^2}{2\sigma^2}} \qquad (4)$$

where $T_0$ is the temperature at which the value of $\alpha(T)$ is optimal and equals $\alpha_{opt}$. $\sigma$ denotes the performance breadth, the temperature range over which the species can reproduce.

The handling time *h(T)* of species is obeying Holling type-II functional response[45] exhibiting a hump or a U-shaped relationship with temperature (Eq. 5), can be represented by an inverted Gaussian function[46].

$$h_i(T) = h_{opt} \cdot e^{\frac{(T-T_0)^2}{2\sigma^2}} \qquad (5)$$

where $h_{opt}$ represents the value of *h(T)* at the optimum temperature $T_0$. $\sigma$ denotes the performance breadth.

The per capita decay rate of pollinators $\kappa(T)$ is observed to follow the Boltzmann-Arrhenius relationship (Eq. 6)[20].

$$\kappa_i(T) = \kappa_{opt} \cdot e^{A_k\left(\frac{1}{T_0} - \frac{1}{T}\right)} \qquad (6)$$

where $\kappa_{opt}$ represents the value of $\kappa(T)$ at the optimum temperature $T_0$. $A_k$ is the Arrhenius constant, which quantifies how fast the decay rate increases with increasing temperature.

Temperature response of the per capita intra-specific coefficient $\beta(T)$ tends to increase monotonically with temperature as given by Boltzmann-Arrhenius relationship (Eq. 7)[17].

$$\beta_{ij}(T) = \beta_{opt} \cdot e^{A_k(\frac{1}{T_0} - \frac{1}{T})} \qquad (7)$$

where $\beta_{opt}$ represents the value $\beta_{ii}(T)$ of at the optimum temperature $T_0$. $\beta_{opt}$ and $A_k$ is different for different regions[20]. Interspecific competition ($i \neq j$) is taken as one-fifth of intraspecific competition[47].

**Simulation**

***Parameter values:*** We obtained experimentally derived thermal tolerance parameters optimum temperature $T_{opt}$ and critical thermal minima $Ct_{min}$ for a set of terrestrial ectotherms (n = 38) published by Deutsch et al[17]. The authors gathered data from 31 thermal performance studies published between 1974 and 2003 based on a collection of insects from 35 different point locations. We calculated the optimal temperature $T_{opt}$ by averaging the overall mean of all regions mentioned in Deutsch et al.[17] specifically within the Köppen climate classification regions of temperate (Cfb), mediterranean (Csa), and tropical (Aw).

Performance breadth ($\sigma$) is calculated using the formula (Eq. 8)[17]

$$\text{Performance breadth}(\sigma) = \frac{T_{opt} - Ct_{min}}{4} \qquad (8)$$

$T_{opt}$ and $\sigma$ are used to predict biological parameter values to CMIP6 temperature (Table S2). The optimum biophysical parameter values (Table S2) e.g., growth rate $\alpha_{opt}$, decay rate $\kappa_{opt}$, handling time $h_{opt}$ are obtained from Jiang et al [29], and, interspecific competition $\beta_{opt}$ from Scranton and Amarasekare [20].

***Initial condition:*** To generate the initial conditions for the historic period (before 1850), we simulated the mutualistic network model with a very low initial value (0.001) using the same parameters and the mean monthly temperature values from 1850 to 2014, based on the assumption that the insect population was low following a large mass extinction event[48]. We then simulated the abundance of each species using the last time step value as the initial population for the historical period (1850–2014), simulated using absolute monthly temperatures. The population state in 2014 was subsequently used as the initial population for the future period simulation (2015–2100).

***Abundance management strategies:*** Abundance management is a strategic approach to species conservation that focuses on maintaining the populations of target species through direct interventions[49]. For abundance management strategies, we have followed 4 strategies: single and multi-pollinator management[50] and single and multi-plant management. Single-species management focuses on the abundance maintenance of a single species through targeted actions like habitat restoration or species protection measures.

Multi-species management, on the other hand, addresses the needs of multiple species simultaneously, considering their interactions and shared habitats, to create more comprehensive and ecosystem-wide conservation strategies[51].

*Without management:* In this simulation, we did not manage any particular species. Instead, we simulated how species abundance will vary with changing monthly temperature from 2015 to 2100 for SSP2-4.5 and SSP5-8.5 scenarios, considering different levels of mutualistic strength, $\gamma_0$. To generate varying mutualistic strength, we used discrete values within the range of mutualistic strength (3 to 0.001)[29]. Monthly population abundance was computed using the population equation of both plants and pollinators, where biological parameters are dependent upon temperature. Then, the annual mean abundance was calculated by averaging the monthly population.

*Single-species management:* Single-species management includes single pollinator management and single plant management. In single-pollinator management, we fixed the abundance of the highest degree pollinator[52]—the pollinator with the maximum interactions with plants in a particular network—to its 2014 population level throughout the simulation period (2015–2100), while all other parameters were simulated as in the previous scenario[29]. Similarly, in single-plant management, we fixed the abundance of the highest degree plant, with all other details following the same approach as for pollinators.

*Multi-species management:* Multi-species management includes both multi-pollinator and multi-plant management. In multi-pollinator management, we fixed the population levels of the top 10% most connected pollinators in each network to their maximum 2014 levels, maintaining these levels throughout the simulation period (2015–2100). All other parameters were simulated as in previous scenarios. Similarly, in multi-plant management, we fixed the top 10% most connected plants to their maximum 2014 levels, with the rest of the simulation details following the same approach as in multi-pollinator management.

**Management strategies' evaluation by mean abundance:** We have calculated the mean annual abundance of pollinators for different mutualistic strengths, ranging from 2015 to 2100. The effect of management strategies on the mean pollinator population is assessed by the Management Efficiency Ratio (MER) (Eq. 9).

$$MER = \frac{A_m - A_w}{A_w} \qquad (9)$$

where $A_m$ is mean abundance under any management, $A_w$ is mean abundance in without management. It is the ratio that states how each management strategy is better than without management. Here $A_m$ means 4 different management strategies, e.g., single-plant management, single-pollinator management, multi-plant management, multi-pollinator management.

**Management strategies' evaluation by diversity and evenness:**

Analyzing diversity and evenness alongside mean abundance offers a comprehensive assessment of ecosystem health by revealing species distribution patterns and the overall balance within the community.

***Rank-abundance curve:*** A rank abundance curve (RAC) is a plot used in ecological studies to display the relative abundance of species in a community, with species ranked from most to least abundant on the x-axis and their relative abundances on the y-axis. It helps visualize species richness and evenness, offering insights into community structure and biodiversity[53,54]. The slope of the RAC illustrates evenness: a steep slope indicates low evenness (few species dominate), while a flatter slope suggests high evenness (similar abundances among species). Diversity is reflected in both the length and shape of the curve. A longer curve indicates higher species richness. The combination of length and slope offers insights into overall diversity, with high diversity characterized by both a high number of species (richness) and a relatively even distribution of individuals among those species (evenness). By using the rank abundance curve, we compared the above-mentioned strategies. But this is only a qualitative interpretation of evenness and abundance. So, we have also used quantitative methods to assess the diversity and evenness of the community, i.e., Shannon diversity and Pielou evenness indices under different management scenarios. The Shannon diversity index assesses diversity based on species richness and abundance, while the Pielou evenness index measures evenness in abundance distribution. These metrics complement RAC insights, enabling more accurate ecological comparisons and analyses.

***Shannon diversity index:*** To measure the diversity of species of different mutualistic strengths, we computed the Shannon diversity index (Eq. 10), which quantifies the diversity of a community by accounting for both the number of different species present and their relative abundances[55,56].

$$H' = -\sum_{i=1}^{n} p_i \ln(p_i) \qquad (10)$$

where *H'* is the Shannon diversity index, *n* is the total number of pollinators and $p_i$ is proportion of pollinators belonging to the *i*-th species of the total number of individuals.

***Pielou evenness index:*** The Pielou Evenness Index quantifies the evenness of species abundance within an ecosystem (Eq. 11). It compares the Shannon Diversity Index (which incorporates both species richness and evenness) to the maximum possible diversity given the number of species present[57]. So, a higher Pielou evenness index indicates a more even distribution of individuals among species, reflecting greater evenness in the ecosystem.

$$J' = \frac{H'}{\ln(n)} \qquad (11)$$

where *J'* is the Pielou evenness index, *H'* is Shannon diversity index and *n* is total no. of pollinators in the community.

**Statistical analyses:** To check whether the pollinator population under each management strategy differs significantly from the population without management, we conducted a two-sample independent t-test. This test assesses whether there is a statistically significant difference between the populations of two distinct groups—in this case, each management strategy compared to the without management. We performed the significance testing at three significance levels: 1% (0.01), 5% (0.05), and 10% (0.1). Since multiple tests were conducted simultaneously, we applied the Bonferroni correction. This correction adjusts the significance levels to account for multiple comparisons, thereby reducing the risk of Type I errors (false positives). By dividing the significance level by the number of tests, we ensure that the overall chance of finding a significant difference due to random variation alone remains within acceptable limits.

**Acknowledgements:**

This research was primarily funded by IIT Gandhinagar. The authors acknowledge the ANRF (SERB) CRG grant (awarded to Udit Bhatia) #CRG/2023/001438 and Pandya Shivpuri chair endowment funds. The authors also thank the members of the Machine Intelligence and Resilience Laboratory at IIT Gandhinagar for their valuable discussions and constructive feedback on this manuscript. Data was obtained from http://www.web-of-life.es, developed by Jordi Bascompte's lab.

**Author contributions:**

Conceptualization: AD, UB, SD, TG, ARG. Analysis: AD, SD. Visualization: AD, UB, SD, TG. Writing – original draft: AD, UB. Writing – review and editing: AD, UB, SD, TG, ARG.

**Competing interests:**

The authors declare that they have no competing interests.

**Data and materials availability:**

Data for 11 mutualistic plant-pollinator networks (Table S1) was obtained from the Web of Life database (www.web-of-life.es). The computer code for analyzing data and creating the plots was written in Python. The code, along with the datasets, are available at https://github.com/AdrijaEco/Ecology_management

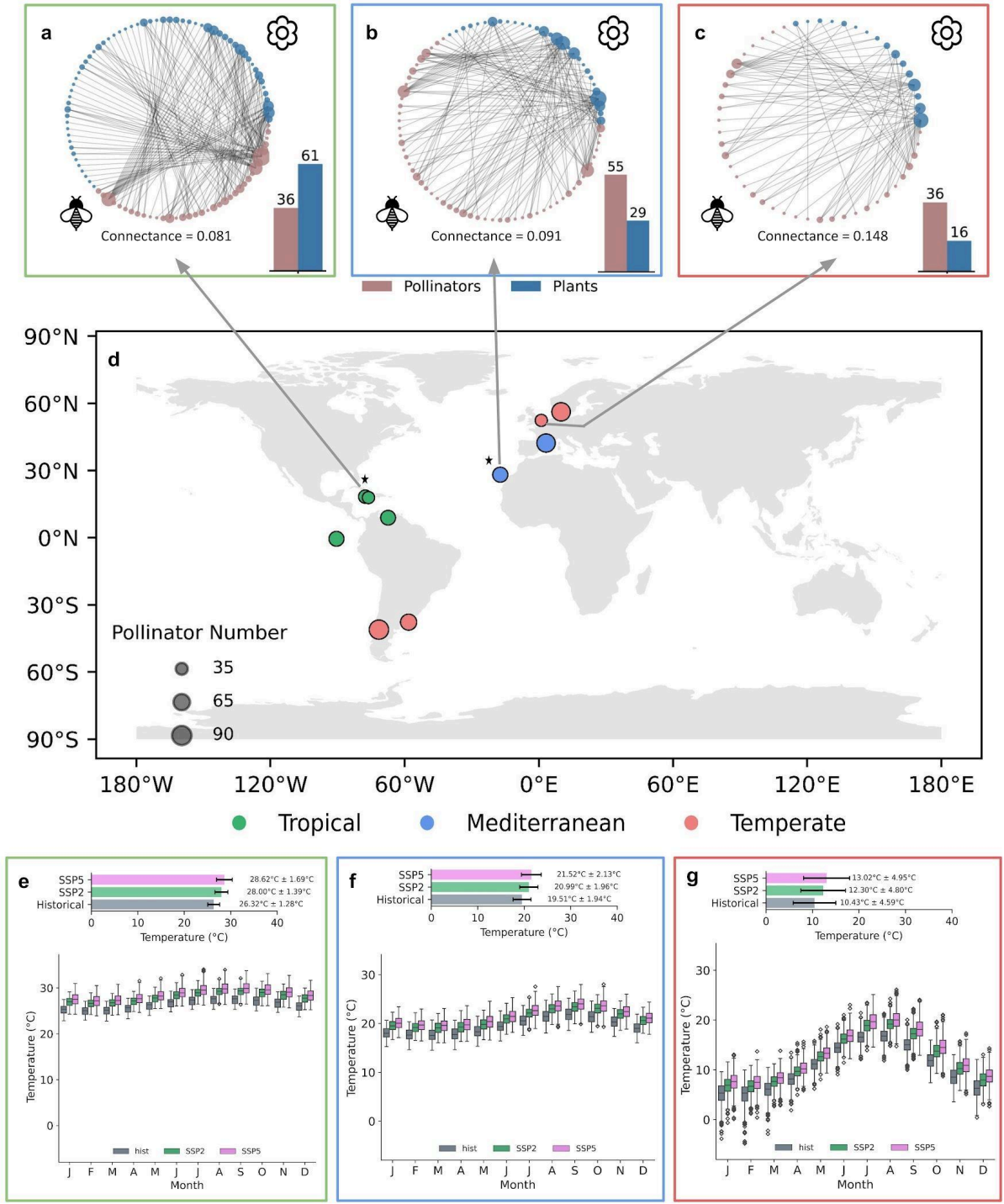

**Figure 1: Plant-pollinator networks and seasonal temperature variation:** Illustrations of plant-pollinator networks in the Tropical **(a)**, Mediterranean **(b)**, and
Temperate **(c)** regions. Node sizes indicate the number of connections for each species. Bar plots indicate the number of species, i.e. plants and pollinators in each network.
**(d)** Geographic locations of 11 terrestrial plant-pollinator networks, with '*' denoting
paired networks. Seasonal temperature variation for networks in the Tropical **(e)**, Mediterranean **(f)**, and Temperate **(g)** regions. The bar plot **(top)** shows the accumulated mean temperature and variability of historical (hist), SSP2-4.5 (SSP2), and SSP5-8.5 (SSP5). Box plot **(bottom)** shows the seasonal temperature variation under historical (1850-2014) SSP2-4.5 and SSP5-8.5 scenarios (2015-2100).

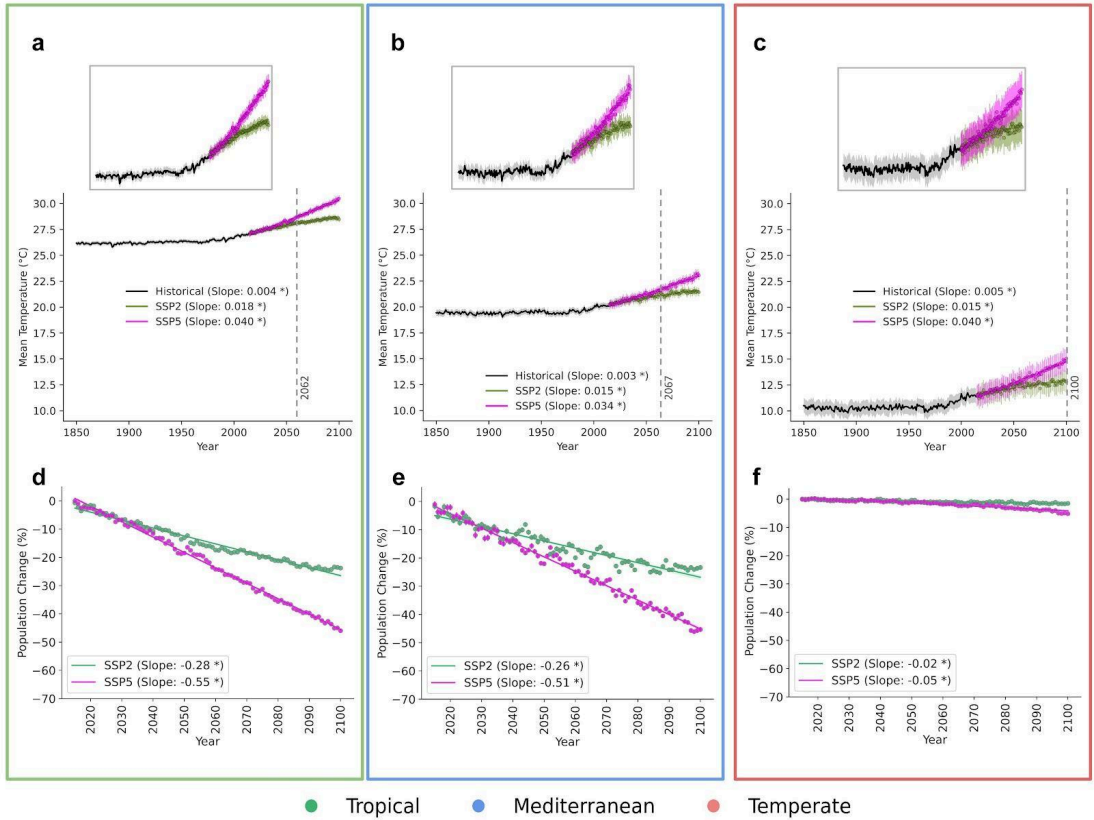

**Figure 2: Population abundance of insect pollinators under climate warming scenarios**: Annual temperature of historical (1850-2014) and SSP2-4.5 and SSP5-8.5 (2015-2100) scenarios in the Tropical **(a)**, Mediterranean **(b)**, and Temperate **(c)** network locations. The vertical line shows the year from which the temperature of SSP2 and SSP5 diverges. Annual mean pollinator population abundance percentage change under warming scenarios SSP2-4.5 and SSP5-8.5 (2015-2100) w.r.t. 2014 population level in moderate environmental health conditions ($γ_0$ = 1.5) in the Tropical **(d)**, Mediterranean **(e)**, and Temperate **(f)** networks, respectively. (*) here slope is significant (at 5% significance level).

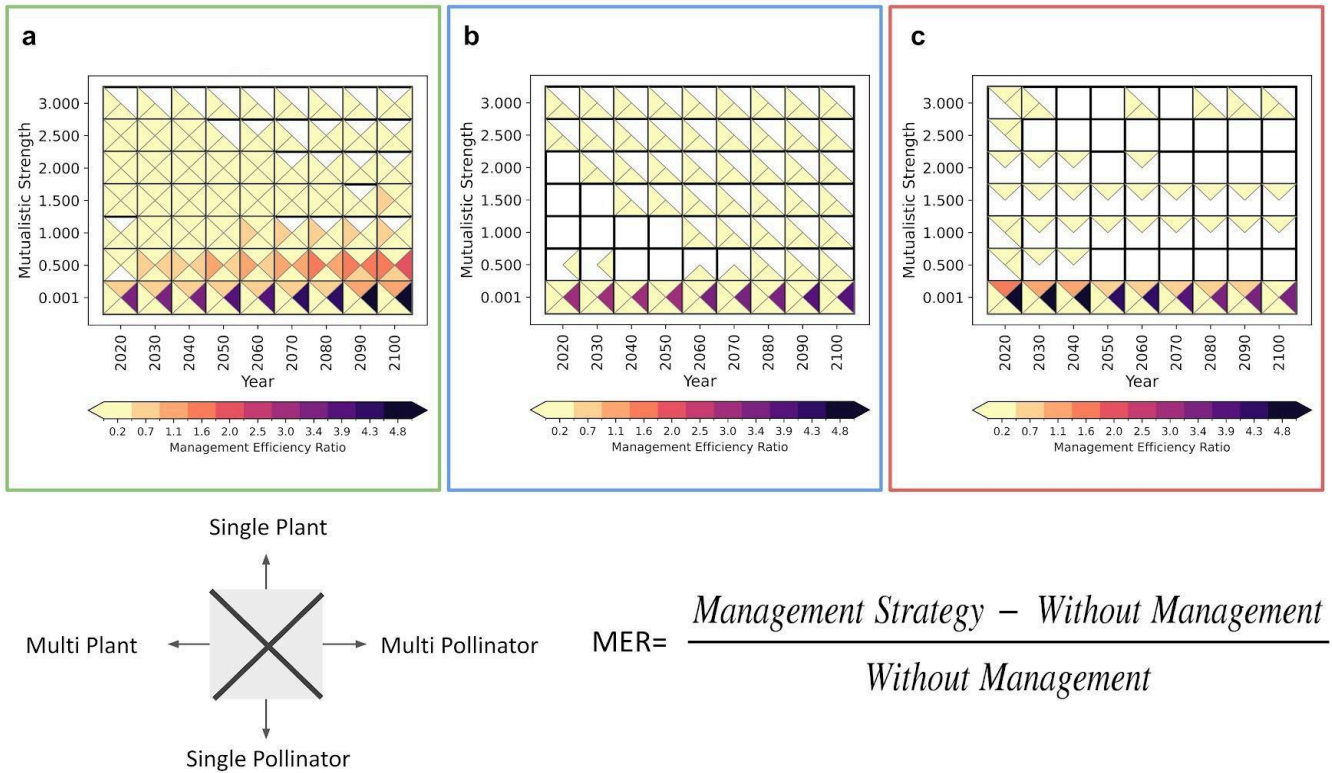

**Figure 3: Comparison of Management Efficiency Ratio across environmental conditions and regions:** Management Efficiency Ratio (MER) of mean pollinator abundance at 10-year intervals with respect to varying mutualistic strength for SSP5-8.5 scenario in the Tropical (a), Mediterranean (b), and Temperate (c) networks. Here, white cells show a value less than equal to 0 means there is no effect of management.

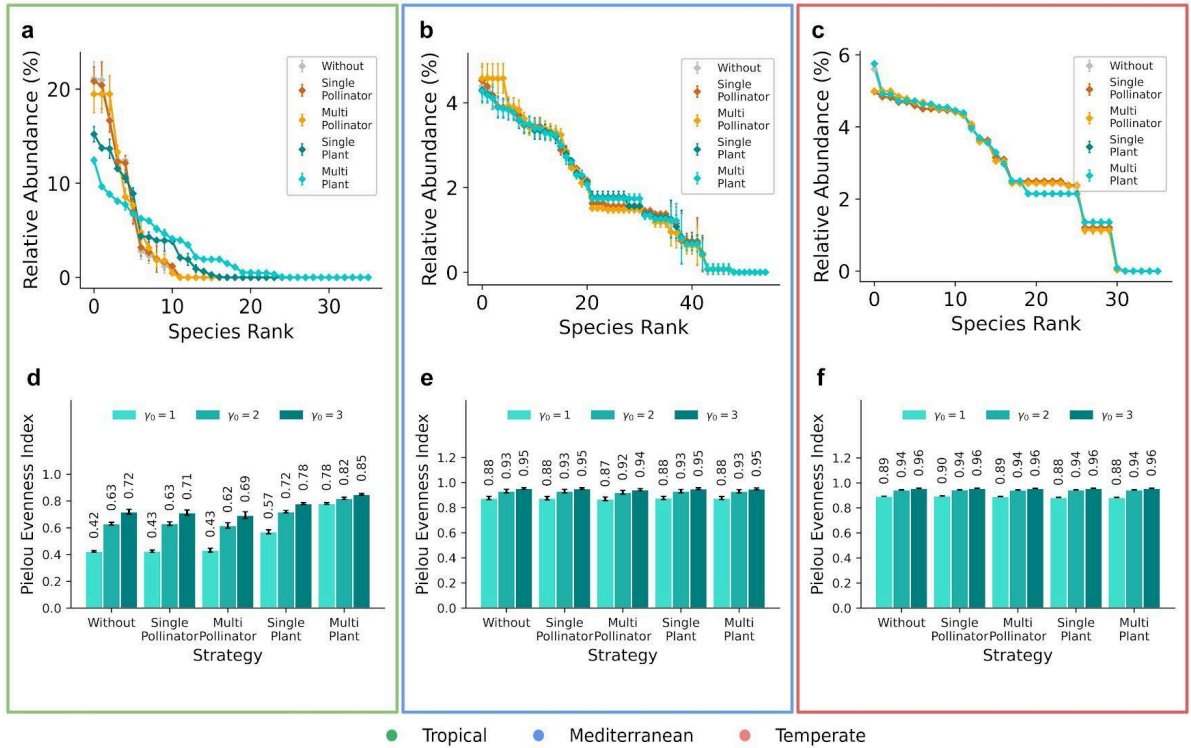

**Figure 4: Comparison of management strategy on species evenness:** Rank abundance curve for pollinators at mutualistic strength ($γ_0 = 1.5$) under SSP5-8.5 in the Tropical (a), Mediterranean (b), and Temperate (c) networks. Variation in the Pielou evenness index under SSP5-8.5 of Tropical (d), Mediterranean (e), and Temperate (f) networks under different mutualistic strengths for different management strategies.